\documentclass[12pt]{iopart}

\def\0{{\sst{(0)}}}
\def\1{{\sst{(1)}}}
\def\2{{\sst{(2)}}}
\def\3{{\sst{(3)}}}
\def\4{{\sst{(4)}}}
\def\5{{\sst{(5)}}}
\def\6{{\sst{(6)}}}
\def\7{{\sst{(7)}}}
\def\8{{\sst{(8)}}}

\def\half{{\textstyle{1\over2}}}

  \let\g=\gamma \let\d=\delta \let\e=\epsilon
  \let\q=\theta  \let\k=\kappa
 \let\m=\mu   
\let\s=\sigma \let\t=\tau    
  \let\D=\Delta  
    
 \let\W=\Omega   \let\G=\Gamma
  
\def\nn{\nonumber} 

\let\pa=\partial  
\newcommand{\be}{\begin{equation}} 
\newcommand{\ee}{\end{equation}}
\def\ba{\begin{array}}
\def\ea{\end{array}}

\def\sst#1{{\scriptscriptstyle #1}}

\newcommand{\bea}{\begin{eqnarray}} 
\newcommand{\eea}{\end{eqnarray}}

\begin{document}

\title{An overview of 
branes in the plane wave background}

\author{Kostas Skenderis\dag\ and Marika Taylor\ddag  
\footnote[3]{skenderi@science.uva.nl and taylor@phys.uu.nl}
}

\address{\dag\  Institute for Theoretical Physics,
University of Amsterdam,
Valckenierstraat 65, 1018XE Amsterdam, The Netherlands}

\address{\ddag\ Spinoza Institute, University of Utrecht,\\
Postbus 80.195, 3508 TD Utrecht, The Netherlands}

\begin{abstract}

We give an overview of D-branes in the maximally supersymmetric
plane wave background of IIB supergravity. We start by reviewing
the results of the probe analysis. We then present the 
open string analysis and show how 
certain spacetime symmetries are restored using worldsheet symmetries.
We discuss the construction of these branes as
boundary states and summarize what is known about the dual gauge theory
description.

\end{abstract}




We present in this contribution an overview of 
D-branes in the maximally supersymmetric plane wave background of 
IIB supergravity 
(henceforth called ``the plane wave''),
\bea  \label{ppwvmet}
ds^2 &=& 2 dx^{+} dx^{-} + \sum_{I=1}^{8} 
( dx^{I} dx^{I} - \mu^2 (x^I)^2 (dx^{+})^2), \\ 
F_{+1234} &=& F_{+5678} = 4 \mu. \nn
\eea
The flux breaks the symmetry of the directions transverse
to the lightcone from $SO(8)$ to $SO(4) \times SO(4)$; we denote this splitting
of the transverse coordinates as $(4,4)$. It is useful to adopt 
the notation $(m,n)$ to indicate the transverse directions wrapped
by a given brane. 

The motivation for studying D-branes in this background 
is twofold. Firstly, the plane wave background is 
related to $AdS_5 \times S^5$ via a Penrose limit \cite{blau}. 
This implies a correspondence between string theory on the 
plane wave background and gauge theory \cite{BMN}.
Secondly, this background is one of the very few RR backgrounds
on which string theory is exactly solvable \cite{Met}.
Given the importance of string theory on such backgrounds, 
it is worthwhile to obtain as thorough an understanding as possible
in this case where the analysis is tractable.

D-branes can be analyzed by a variety of methods.
In string perturbation theory they are surfaces on which strings 
can end and thus one can investigate possible D-branes by 
finding consistent boundary conditions for open strings 
on the background (\ref{ppwvmet}). In the Green-Schwarz (GS)
formulation the spacetime symmetries (the symmetries of the 
D-brane) manifest themselves as global symmetries of the worldsheet action.

Conformal invariance of the open string requires that 
the coordinates that specify the position of the 
brane, the worldvolume gauge field and the associated fermions
satisfy the Dirac-Born-Infeld (DBI) field equations. 
One may thus obtain possible D-branes by solving
the DBI field equations. The target space symmetries
of the GS open string should appear as worldvolume symmetries 
of the DBI action.

D-branes can also be described using boundary states.
The boundary state is a closed string state that imposes  
appropriate boundary conditions. Open-closed duality  
requires that D-branes have such a description and one may thus 
investigate possible D-branes by constructing boundary states. 
The symmetries that the boundary state preserves are the 
combinations of the closed string symmetries that 
leave the boundary state invariant.

In the next three sections we will discuss these three complimentary
approaches; the results are summarised in the tables 1 and 2.
Plane wave/gauge theory duality 
implies that the D-branes should have a gauge theory description and
this will be discussed in section 4, followed by conclusions in
section 5. The results described here were obtained in \cite{ST1,ST2,ST3},
with related work in  \cite{boundary,open,dual,giant}. 
The emphasis in this contribution
will be on the results and on the methods used rather on the technical details. 

\begin{table}
\caption{\label{perD} D-branes visible in string perturbation theory
and their supersymmetries. $(m,n)$ denotes 
the number of worldvolume coordinates wrapping the
two sets of directions transverse to the lightcone. 
The branes in the open string channel also wrap 
the lightcone, but the corresponding boundary states
do not. $q^-$ denotes dynamical 
supersymmetries and $q^+$ kinematical supersymmetries. The 
hatted quantities are symmetries that were restored using 
worldsheet symmetries. $x_0$ denotes the constant position of the brane in  
transverse space.}
\begin{indented}
\item[]\begin{tabular}{@{}llll}
\br
Branes & Probe analysis & Open string analysis & Boundary States \\ 
\mr
$D_-$ & & & \\
\mr
$(m,m \pm 2)$ at $x_0=0$ &  8 $q^-$ + 8 $q^+$ & 8 $q^-$ + 8 $q^+$ & 
8 $q^-$ + 8 $q^+$\\
$(m,m \pm 2)$ at $x_0 \neq 0$ &  \qquad \quad 8 $q^+$ & 8 $\hat{q}^-$ + 8 $q^+$ & 
8 $q^-$ + 8 $q^+$\\
\mr 
$D_+$ & & & \\
\mr 
$(0,0)$ at $x_0$  & 8 $q^-$ &  
8 $q^-$ + 8 $\hat{q}^+$ & 8 $q^-$ + 8 $q^+$ \\
$(0,4)$ with flux & 8 $q^-$ & 8 $q^-$ + 8 $\hat{q}^+$ & 8 $q^-$ + 8 $q^+$ \\
$(m,n)$ at $x_0$  & 0 &  
\qquad \quad 8 $\hat{q}^+$ & \qquad \quad 8 $q^+$ \\
\br
\end{tabular}
\end{indented}
\end{table}

\begin{table}
\caption{\label{nonperD} Branes that wrap only one of the 
lightcone directions. These branes are not visible in the 
string perturbation theory in the lightcone gauge.}
\begin{indented}
\item[]\begin{tabular}{@{}ll}
\br
Branes & SUSY \\ 
\mr
$(+,0,S^3)$ and $(+,S^3,0)$ & 16 $q^-$ \\
$(+,0,1)$ and $(+,1,0)$ & 8 $q^-$ \\
\br
\end{tabular}
\end{indented}
\end{table}

\section{Probe analysis}

As discussed, one may investigate possible D-branes by 
looking for solutions of the DBI field equations. The gauge invariant 
DBI field equations for bosonic fields were derived in generality 
in \cite{ST1}. These equations can be effectively used
to look for possible D-brane embeddings.  
In \cite{ST1} we systematically analyzed embeddings with constant transverse 
scalars and zero worldvolume flux and furthermore 
discussed some specific cases with non-zero fluxes. 

The DBI action is invariant under local kappa symmetry 
and target space supersymmetries. The combined 
transformations of the (target space) fermions are 
\be \label{kappa}
\delta \theta = (1 + \Gamma) \kappa + \e
\ee
where $\G$ is the $\k$ symmetry projector and 
$\e$ is the target space Killing spinor evaluated on the 
worldvolume. The worldvolume supersymmetry arises after gauge 
fixing the kappa symmetry. A convenient gauge fixing is the one 
that uses the kappa symmetry projector evaluated on the 
embedding, $(1 - \Gamma) \theta =0$  \cite{kappa}.
Combining this projection with (\ref{kappa}) we get 
the condition for unbroken supersymmetry
\be \label{susy}
\G \e = \e
\ee
where both the kappa symmetry projector $\G$ and the Killing 
spinor $\e$ are evaluated on the brane.

In the generic case, the solutions of (\ref{susy}) give the number of 
supersymmetries linearly realised on the brane. 
The open string analysis discussed in the next section, however,
suggests that the DBI action on the plane wave background has additional 
supersymmetries. It is an interesting problem to analyze whether 
there are special class of backgrounds on which the DBI action exhibits 
more symmetries than those implied by the generic analysis.

The presence of a brane breaks some of the target space isometries
and one can obtain new embeddings by acting with the broken generators.
Since all properties of the new brane follow from those 
of the original brane, one may group the branes in equivalence 
classes: two branes are considered equivalent if they are related 
by the action of a broken isometry of the background. We call such
branes ``symmetry-related branes''. Since the translational 
Killing vectors in the plane wave are $x^+$ dependent, the same equivalence class 
may contain both static and time-dependent branes. In particular,
static branes localized at the origin 
and certain time-dependent branes localized away from origin belong to 
same equivalence class. Notice also that from the point of view
of open strings a brane localized at the origin and one
located at a constant (non-zero) position are not symmetry-related.

The supersymmetric embeddings found are summarized in tables 1 and 2.
The kappa symmetry analysis shows that the preserved supersymmetry of the 
brane 
depends both on the splitting $(m,n)$ and on the transverse positions.
For instance $(m,m \pm 2)$ branes, which we denote $D_-$ branes, 
preserve 16 supercharges
when located at the origin in the transverse space, but appear to 
preserve only the 8 kinematical supercharges\footnote{
We call kinematical the supercharges that anticommute to the lightcone 
momentum and dynamical the ones that anticommute to the lightcone
Hamiltonian, possibly along with other generators.}
when located away from the 
origin. Switching on constant worldvolume fluxes
may in some cases allow the brane to be at 
specific non-zero position whilst preserving 16 supercharges \cite{ST1}.
  
In the plane wave one can also have branes wrapping only one lightcone
direction, $x^+$. We should emphasise that these have non-degenerate induced
metrics, cf corresponding null branes in flat space.  
Only specific embeddings seem to preserve any
supersymmetry and these are listed in Table 2: there are D-strings 
and D3-branes which wrap arbitrary radius three spheres. The explicitly
broken translational invariance in the $x^-$ direction means that the
lightcone momentum $p^+$ is not preserved, and thus neither are kinematical
supercharges which anticommute to $p^+$. 

Finally, one has instantonic branes which do not wrap the lightcone.
These are described by closed string boundary states and are related
to the branes wrapping the lightcone under open/closed duality. We will
return to this subject later. 

\section{Open string analysis}

In string perturbation theory D-branes are surfaces on which strings can 
end. To classify D-branes one may thus investigate consistent 
open string boundary conditions. At tree-level this amounts to
requiring that the variational problem is well defined, i.e. requiring that 
$\d S=0$ implies the field equations. For a theory defined on a 
manifold (worldsheet)
with a boundary this requires that boundary conditions are chosen such that 
all boundary terms arising in the variation are zero. Further conditions
may arise at the quantum level. 
 
Such an analysis of open strings in the plane wave background
was presented in \cite{ST2}. One finds that 
one can impose standard static D-brane boundary conditions
on the bosonic coordinates, but also that 
more general time-dependent 
boundary conditions are possible. For instance, the ends of the string may 
trace a trajectory of 
an ellipse with angular velocity equal to the mass 
of the string. 
This is an example of a symmetry-related brane
though not all time dependent branes we find
are symmetry-related.

Appropriate boundary conditions for the fermions $(\q^1,\q^2)$ amount to 
setting to zero on the worldsheet boundary a specific linear combination 
of them,
$(\q^1 - \W \q^2)|=0$, where $\W$ is a $16 \times 16$ dimensional 
matrix. We find that D-branes have different properties depending 
on whether $(\W \Pi)^2 =-1$, where $\Pi=\g^1 \g^2 \g^3 \g^4$, 
or $(\W \Pi)^2 =1$. We call the former branes $D_-$ branes and the former
$D_+$ branes; these two categories reflect the splitting $(m,n)$ as indicated in table 1. 

After the D-brane has been defined, we can investigate its symmetries
by looking for symmetries of the worldsheet action with these boundary
conditions. Generically the symmetries of the open string are the
closed string symmetries that respect the boundary conditions. 
The supersymmetries obtained in this fashion agree exactly 
with the ones obtained in the probe analysis using (\ref{susy}).

In some circumstances, however, the worldsheet theory may 
admit additional symmetries. This is the case for the open string 
on the plane wave. The (gauge-fixed) worldsheet action in this case is that 
of free massive bosons and fermions. The fact that 
the Lagrangian is quadratic in the fields implies that it 
is invariant up to a total derivative under a transformation
that is a shift of the fields by a parameter that satisfies the free 
field equations. Let us illustrate this point with a free 
massive boson; an exactly analogous discussion holds for free 
massive fermions. Consider the action
$
S = \int d \t \int_0^\pi d\s \half (\pa^\m X \pa_\m X + m^2 X^2).
$
Under the transformation 
\be \label{tra}
\d X = \e(\s,\t) 
\ee
we obtain
\be \label{var}
\d S = \int d \t d\s X (\nabla^2 - m^2) \e 
+ \int d \t [(\pa_\s \e) X ]_{\s=0}^{\s=\pi}.
\ee
It follows that if $\e(\s,\t)$ satisfies the free field equation the
variation results in only a boundary term. 

If in addition $\e(\s,\t)$ satisfies appropriate boundary conditions so 
that the boundary term vanishes we obtain a new symmetry. Expanding 
$\e(\s,\t)$ in a basis we obtain in this way a countably infinite number of symmetries.
The corresponding Noether charges evaluated on-shell are linear in the 
oscillators. These symmetries reflect the fact that the mode expansion 
for the quantum field can be solved explicitly. They are a generalization
of the symmetries generated by the chiral currents, $J_n = z^n \pa X$, 
in the case of CFTs associated with free massless bosons. 

Since the open string worldsheet Lagrangian is the same 
as the closed string Lagrangian, the variation 
of the open string action under any symmetry transformation of the 
closed string either leaves the action invariant or results in a 
boundary term. In the latter case one would conclude that this
symmetry is broken. We have just seen, however, that the 
open string action also varies into a boundary term under a 
transformation of the form (\ref{tra}). It follows that if 
we can find appropriate $\e(\s,\t)$ such that the boundary term
in (\ref{var}) cancels the boundary term in the variation 
of the action under the closed string transformation, then 
the combined worldsheet and closed string transformation 
is a good symmetry of the open string. We will denote the
generators of these new symmetries by the same symbol
as the generator of the corresponding closed string symmetry, 
but with a hat; this notation is used in table 1. 

We have shown in \cite{ST2,ST3} that one can use this mechanism 
in order to restore some apparently broken symmetries. 
In all such cases the violation 
of the closed string symmetry depends solely on quantities
that are determined by the boundary conditions, and the 
violating terms can be adjusted to zero by changing
the boundary condition.  For example, $D_-$ branes 
located at a constant transverse position $x_0$
appear to break all dynamical supersymmetries, and 
the violating terms vanish when the transverse position
is set to zero, $x_0=0$. We find that in these cases 
one can combine a closed string transformation with a
transformation of the form (\ref{tra}) (and a corresponding
fermionic transformation) to obtain a good symmetry of the 
open string. 

In all these cases the brane located at the origin is symmetric 
without the need of extra worldsheet symmetries. 
The Noether charge for this symmetry is exactly equal on-shell to 
the Noether charge that generates the new symmetries for the brane
located away from the origin. Taking into account the extra symmetries
one finds that all on-shell conserved charges are exactly the 
same for static branes located at and away from the origin. 
The only exception is the lightcone Hamiltonian. The 
lightcone Hamiltonian for a brane located at $x_0$  
contains an extra c-number contribution $\D H$ that is equal to the 
energy that a classical open string with ends at $x_0$ has due 
to the harmonic oscillator potential. $\D H$ itself
is the on-shell value of a worldsheet charge. These considerations
imply that the superalgebras for the brane at and away from the origin
are almost identical. The only differences result from the difference
in lightcone Hamiltonians.

Similar considerations lead to new kinematical symmetries
in the cases where the closed string kinematical supersymmetries
are not compatible with the boundary conditions.
The fact that the worldsheet 
theory contains free fermions always implies that there is 
a corresponding kinematical supercharge: it is given 
on-shell by the zero mode of the free fermion.

The spectrum of the theory reflects the existence of all the 
symmetries discussed here. In particular, the states 
of the theory organize themselves into supermultiplets
of the extra supersymmetries. For the $D_-$ branes the 
theory has a unique vacuum state. The dynamical supersymmetries
commute with the lightcone Hamiltonian, and thus its eigenstates
form multiplets under the dynamical supersymmetry. The kinematical 
supercharges are spectrum generating. For the $D_+$ branes
both the dynamical and the kinematical supercharges commute with the 
Hamiltonian and the ground state is degenerate.

\section{Boundary states}

Boundary states in the plane
wave background were discussed in \cite{boundary,ST3}.
The boundary state is a closed string state
that represents the addition of a boundary 
to the tree level worldsheet. It is 
constructed by imposing the boundary conditions
on the worldsheet fields as operator relations.
The appropriate boundary is now spacelike: 
at time $\t=\t_0$ a closed string is created (or
annihilated) from the vacuum. 
The gluing conditions can be 
solved to obtain the boundary state as a coherent 
state of closed strings.

Having obtained the boundary state one may then check what 
symmetries it preserves. This can be done using the 
explicit expressions for the closed string generators
in terms of modes. Acting on the boundary state and using the 
gluing conditions one can uniquely determine which of the 
closed string symmetries are preserved by the boundary state.

In all cases we find \cite{ST3} 
that the symmetries of the corresponding boundary 
state match the symmetries of the brane in the 
open string channel. However, one does not use any additional 
worldsheet transformations.
In particular, in all cases the boundary states
are annihilated by a combination of the closed 
string kinematical supercharge generators. Furthermore, the boundary 
state representing 
a $D_-$ brane located away from the origin preserves 8 closed string 
kinematical supercharges, and 8 linear combinations of the 
the closed string dynamical and kinematical supercharges.

\section{Dual gauge theory}

The branes whose dual description is best understood are the 
$D_-$ branes \cite{dual,ST1}. The idea behind the construction is to 
add a $Dp$ brane in the sequence:
\be
D3 \quad \stackrel{near-horizon}{\longrightarrow} AdS_5 \times S^5 \quad
\quad \stackrel{Penrose}{\longrightarrow} \quad {\rm plane\ wave}
\ee
We have shown in \cite{ST1} that the $D_-$ branes localized 
at the origin originate 
from half supersymmetric $AdS$ embeddings in $AdS_5 \times S^5$ 
in the Penrose limit.
The latter in turn arise from supersymmetric intersections of the 
$Dp$ branes with $D3$ branes in the near horizon D3-brane limit.
This sequence of relations allows one to identify 
the dual description of $D_-$ branes as being represented by 
a defect CFT. The latter captures the physics of the $3-p$ strings
in the original $D3$-$Dp$ system. 

In the BMN proposal \cite{BMN} the closed string ground 
state is constructed in the gauge theory by the operator
${\rm Tr} Z^J$, where $Z=X^1 + i X^2$ and $X^i,i=1,..6$ 
are the six scalars of ${\cal N}=4$ SYM. The $Z$'s represent the 
string bits and by taking the trace one obtains a closed string.
To construct an open string we should not trace over the $Z$'s but 
instead put ``quarks'' at the end of the strings. These 
``quarks'' are supplied by the defect theory: they 
are some of the massless modes in the spectrum of the $3-p$ strings.  
The open string 
vacuum is then represented by the defect operator $\bar{q} Z^J q$,
where the $q$'s are the scalars of the hypermultiplet
in the case of ND=4 intersections, but are instead fermions
for ND=8 intersections. In all cases one finds that 
this state indeed has the correct lightcone energy to represent the 
lightcone vacuum. We refer to \cite{ST1} for more details. 

In table 3 we summarize what is known about the relations between 
branes in the plane wave, branes in $AdS_5 \times S^5$ and
brane intersections in asymptotically flat spacetimes. These relations
are instrumental in obtaining the
dual gauge description of branes in the plane wave/gauge theory correspondence.
The first two entries were discussed above.
The $(+,-,0,0)$ D-string is related to the D-instanton via 
open-closed duality and as such may be related to instantons 
in the gauge theory.
The $(+,-,0,4)$ branes most likely represent the Penrose limit 
of a baryon vertex (but they are not the Penrose limit of the 
known baryon vertex solution \cite{mateos}). Giant gravitons 
have also been discussed 
in \cite{giant},
and the $(+,0,1)$ branes are expected to be related to 
monopoles. 

\begin{table}
\caption{\label{emb} Supersymmetric intersections in asymptotically flat 
spacetimes, supersymmetric embedding in $AdS_5 \times S^5$, branes
in the plane wave and the dual description. 
$(r|Dp\perp D3)$ denotes an orthogonal intersection of a $Dp$ brane 
with a $D3$ brane over an $r$-brane.$(+,-,m,n)_{x_0}$ denotes a brane
localized at $x_0$.}
\begin{indented}
\item[]\begin{tabular}{@{}llll}
\br
Intersection & Embedding & brane & dual description\\ 
\mr
$(n| D(2n{+}1){\perp}D3)$ & $AdS_{n{+}2} \times S^{n}$   & 
$(+,-,n{+}1,n{-}1)_{x_0=0}$
& $(n{+}1)d$ dCFT \\
$(n| D(2n{+}5){\perp}D3)$ & $AdS_{n{+}2} \times S^{n{+}4}$   & 
$(+,-,n{+}1,n{+}3)_{x_0=0}$ & $(n{+}1)d$ dCFT \\
\qquad -- &  $R \times S^1$ & $(+,-,0,0)$ & instantons ? \\
\qquad -- &  $R \times S^3$ & $(+,-,0,2)$ & giant graviton ? \\
\qquad --  & ?  & $(+,-,0,4)$ with flux       &  baryon vertex ?                   \\
\qquad --   
& giant graviton & $(+,0,S^3)$ and  $(+,S^3,0)$   &  giant graviton?                   \\
\qquad  -- & rotating $(A)dS_2$ string & $(+,1,0)$ and $(+,0,1)$   
& monopoles ? \\
\br
\end{tabular}
\end{indented}
\end{table}

\section{Discussion}

We presented an overview of branes in the plane wave background.
Perhaps the most surprising result is the restoration of certain
symmetries using worldsheet symmetries. These results were obtained at 
tree level and an important question is whether string interactions
will respect them. Notice that the symmetries of the open string 
match the symmetries of the boundary states only after the new symmetries
are included. Furthermore, the symmetries of the boundary states
are less subtle as they are always a proper subset of the closed string 
symmetries. These considerations suggest that the new symmetries are 
compatible with string interactions.  If this is so, one should 
able to present arguments for the restoration of the symmetries
that are valid for any worldsheet and also one should be able to 
find the extra worldvolume supersymmetries of the DBI action 
in the plane wave background.

\end{document}